\documentclass[]{article}
\usepackage{lmodern}
\usepackage{amssymb,amsmath}
\usepackage{ifxetex,ifluatex}
\usepackage{fixltx2e} 
\ifnum 0\ifxetex 1\fi\ifluatex 1\fi=0 
  \usepackage[T1]{fontenc}
  \usepackage[utf8]{inputenc}
\else 
  \ifxetex
    \usepackage{mathspec}
    \usepackage{xltxtra,xunicode}
  \else
    \usepackage{fontspec}
  \fi
  \defaultfontfeatures{Mapping=tex-text,Scale=MatchLowercase}
  
\fi
\IfFileExists{upquote.sty}{\usepackage{upquote}}{}
\IfFileExists{microtype.sty}{%
\usepackage{microtype}
\UseMicrotypeSet[protrusion]{basicmath} 
}{}
\usepackage[a4paper]{geometry}
\usepackage{longtable,booktabs}
\ifxetex
  \usepackage[setpagesize=false, 
              unicode=false, 
              xetex]{hyperref}
\else
  \usepackage[unicode=true]{hyperref}
\fi
\hypersetup{breaklinks=true,
            bookmarks=true,
            pdfauthor={Yoav Ram, Ofer Moshaioff, Idan Cohen, Omri Dor},
            pdftitle={Forecasting the Israeli 2015 elections using a smartphone application},
            colorlinks=true,
            citecolor=blue,
            urlcolor=blue,
            linkcolor=magenta,
            pdfborder={0 0 0}}
\title{Forecasting the Israeli 2015 elections using a smartphone application}
\author{Yoav Ram, Ofer Moshaioff, Idan Cohen, Omri Dor}
\date{16 March 2015}

\begin{document}
\maketitle

\paragraph{Author details}\label{author-details}

\begin{itemize}
\itemsep1pt\parskip0pt\parsep0pt
\item
  YR: Department of Molecular Biology and Ecology of Plants, Tel-Aviv
  University. Email: yoavram@gmail.com.
\item
  OM: Business School, Tel-Aviv University
\item
  IC: HP Software
\item
  OD: Computer Science Department, Hebrew University of Jerusalem
\end{itemize}

\section{Abstract}\label{abstract}

We developed a smartphone application, Ha'Midgam, to poll and forecast
the results of the 2015 Israeli elections. The application was
downloaded by over 7,500 people. We present the method used to control
bias in our sample and our forecasts. We discuss limitations of our
approach and suggest possible solutions to control bias in similar
applications.

\section{Introduction}\label{introduction}

The 19th Knesset, elected on January 22nd, 2013, was officially
dispersed on December 8th, 2014. The elections for the 20th Knesset,
which were supposed to be held on Noveber 7th, 2017, are to be held on
March 17th, 2015, more than two years before scheduled and just two
years and two months after the previous elections.

During the weeks after the elections were declared, Ofer Moshaioff, Yoav
Ram and Idan Cohen developed a smartphone application (`app') called
Ha'Midgam (\url{http://hamidgam.com}). This app allowed users to
anonymously vote for one of the major participating parties in the
upcoming elections (2015), to disclose their vote in the previous
elections (2013), and to view a forecast of the 2015 election results
based on the aggregated data from all users.

The app was published for Android devices on the
\href{https://play.google.com/store/apps/details?id=com.bmi.midgam}{Android
Play Store} on December 29th, 2014 and for iOS devices (iPhone, iPad;
developed by Elad Ben-Israel) on the
\href{https://itunes.apple.com/il/app/hmdgm/id956943031?mt=8}{Apple App
Store} on January 26th, 2015. It quickly gained media attention on local
radio shows, digital media and newspapers. This media attention
contributed to over 7,500 application downloads by March 16th, 2015.

Our app differs from traditional polls in several aspects. In
traditional polls, media outlets publish forecasts based on a group of
500-1,000 individuals that were chosen by a polling company at a
specific point in time to reflect an unbiased sample of the population.

In contrast, our app allows users to view a realtime, online forecast of
the elections based on individuals that chose to disclose their vote.
Therefore, the sample size in our app is roughly 10-fold. However, in
contrast to traditional polls, our app doesn't collect any demographic
information, such as age, socio-economical status, religion or
ethnicity. Therefore, our app's sample may be biased and therefore
requires statistical manipulation.

Our app does collect information that is unique: the app allows users to
change their mind at any time; it keep a history of user choices; it
logs the precise time and, if allowed by the device, location; and most
importantly for the sake of this manuscript, the app asks users to
disclose which party they voted for in the previous elections (2013).
Our hypothesis was that this information could be enough to make a good
forecast of the elections results - the distribution of seats between
the participating parties.

In this manuscript we describe how the app works, the methods we used to
manipulate the data, and the forecasts we got. We wanted to make this
manuscript available before the elections day begins and therefore this
manuscript in it's current form includes only basic analysis.

\section{Methods}\label{methods}

\subsection{App technical description}\label{app-technical-description}

The mobile client was developed for the Android and iOS smartphone
operating systems (the iOS version didn't include the entire feature
set). The app communicated with a RESTful API server, developed using
\href{http://www.python.org}{Python 2.7} and the
\href{http://flask.pocoo.org/}{Flask} web application framework and
hosted on \href{http://www.heroku.com}{heroku}, largely following a
tutorial by
\href{http://blog.miguelgrinberg.com/post/designing-a-restful-api-with-python-and-flask}{Miguel
Grinberg}.

The app presents to the user a grid of the parties, including some basic
information and a link to the party home or Facebook page. The user can
vote to a specific party, at which point the results forecast screen
appears. The user can view the number of seats per party. At any time
the user can change his vote. In the Android version additional features
were implemented; most importantly, users were asked to disclose their
vote in the 2013 elections. In addition, users could see the a
geographical distribution of the votes by the country main
administrative regions.

\subsection{Seats distribution
forecasting}\label{seats-distribution-forecasting}

We only describe our latest approach with some variations. The basic
problem is how to control bias in our vote sample. Although our sample
has over 7,500 votes, it could be biased due to several factors such as
age, socio-economical status, and party activist propaganda.

\subsubsection{Bias control}\label{bias-control}

We started asking users for their 2013 elections choices on February
13th 2015. We used this information, together with the 2013 elections
\href{http://www.votes-19.gov.il/nationalresults}{official results} to
attempt to control sample bias.

First, we take only the latest vote for each device id, both from the
2013 and the 2015 datasets. Next, we calculate a counts matrix \(C\)
with rows for 2015 parties, columns for 2013 parties: \(C_{i,j}\) is the
number of individuals who voted for party \(j\) in 2013 and will vote
for party \(i\) in 2015.

Next, we use the counts matrix \(C\) to estimate the transition matrix
\(M\) in which \(M_{i,j}\) is the probability that an individual who
voted for party \(j\) in 2013 will vote for party \(i\) in 2015. This
was done by normalizing the columns:
\(M_{i,j} = \frac{C_{i_j}}{\sum_{i}{C_{i,j}}}\).

We then generate the 2013 results vector \(v\) from the official results
data, removing counts of parties for which we have no information as
well as illegal or discarded votes. We multiply the transition matrix by
the results vector to get the forecast vector: \(f = C \cdot v\). The
forecast vector \(f\) describes our prediction of the number of votes
each party will get in the 2015 elections.

To get a forecast of the number of seats for each party we process the
forecast vector \(f\) using the
\href{https://www.knesset.gov.il/lexicon/eng/seats_eng.htm}{Bader-Offer
method}, also known as the
\href{http://en.wikipedia.org/wiki/Hagenbach-Bischoff_system}{Hagenbach-Bischoff
system}. In our version of the Bader-Offer method we disregarded surplus
vote agreements.

The multiplication of the transition matrix with the 2013 results vector
can be viewed as giving different respondents different weights. If
\(c_i\) respondents replied that they voted for party \(i\) in 2013,
then in the normalized transition matrix, each such respondent has
weight \(1/c_i\). When we right-multiply the matrix with the actual 2013
results vector \(v\), each respondent ends up with the weight
\(v_i/c_i\). With this weighting scheme, the total weight of respondents
that claims to have voted for party \(i\) in 2013 is the same as the
actual number of voters for \(i\) in 2013. Our sample now `agrees' with
the actual 2013 election results. To recap, we inserted a weighting
scheme that controlled for the publicly known 2013 election results.

\subsubsection{Additional bias control}\label{additional-bias-control}

As another layer of bias correction, we experimented with fixing the
number of votes received by parties that represent four demographies to
the number of votes in 2013. These demographies are:

\begin{enumerate}
\def\labelenumi{\arabic{enumi}.}
\itemsep1pt\parskip0pt\parsep0pt
\item
  The arab sector, represented by Hadash, Balad \& Raam-Taal in 2013 and
  by the Arab Unified List in 2015.
\item
  The Ashkenazi-Orhodox sector, represented by Yahadut Ha'Tora both in
  2013 and in 2015.
\item
  The Sfaradi-Orthodox sector, represented by Shash and Am Shalem in
  2013 and by Shas and Yachad in 2015. Because Yachad merged with Ozma
  La'Am for the 2015 elections, we includied Ozma La'Am in the
  respective 2013 votes.
\item
  The liberal, pro-cannabis legalisation party, Ale Yarok.
\end{enumerate}

Fixing the number of voters of the first three demographies can be
justified by the relatively constant number of seats their respective
parties received in the previous three elections and by the sectoriality
of these parties. As for fixing the number of votes of Ale Yarok, this
was considered necessary because supporters of this party are known to
be very active online, thus generating biases in online surveys and
polls. For example, the number of ``Likes'' Ale Yarok has in Facebook is
85,709, compared with 27,205 Ha'Likud, the major right winged party,
has.

\section{Results}\label{results}

Using the procedure above and based on the votes as of 16th March 2015,
the app has made these forecasts:

\begin{longtable}[c]{@{}llllll@{}}
\toprule
Party & Raw & Stand. & Fixed: AY & AY, YH, AU & AY, YH, AU,
S\tabularnewline
\midrule
\endhead
Ha'Mahane Ha'Zioni & 32 & 25 & 26 & 25 & 26\tabularnewline
Ha'Likud & 17 & 21 & 22 & 21 & 22\tabularnewline
Yesh Atid & 16 & 15 & 16 & 15 & 16\tabularnewline
Ha'Bayit Ha'Yehudi & 12 & 13 & 14 & 13 & 14\tabularnewline
Yachad & 10 & 9 & 10 & 9 & 9\tabularnewline
Merez & 13 & 9 & 9 & 9 & 9\tabularnewline
Arab Union & 0 & 7 & 8 & 11 & 12\tabularnewline
Kulanu & 6 & 6 & 6 & 6 & 6\tabularnewline
Ale Yarok & 10 & 6 & 0 & 0 & 0\tabularnewline
Shas & 4 & 5 & 5 & 5 & 0\tabularnewline
Yahadut Ha'Tora & 0 & 4 & 4 & 6 & 6\tabularnewline
Israel Beytenu & 0 & 0 & 0 & 0 & 0\tabularnewline
\bottomrule
\end{longtable}

The numbers in the table represent a forecast of the number of seats for
each party. Raw: based on raw data, 7,506 votes. Stand.: data
standardized using a 2013 to 2015 transition matrix (see Methods), 2,447
votes. Fixed: data standardized using the transition matrix, and number
of votes of specific parties were fixed at their 2013 values (AY: Ale
Yarok; YH: Yahaudt Ha'Tora; AU: Arab Union; S: Sash). Note that the
minimal number of seats in the 2015 elections is four. This is in
contrast to previous elections in which the minimal number of seats was
two.

The raw data, with obfuscated timestamps and locations, may be available
upon request to the first author, depending on the purpose of use.

\section{Discussion}\label{discussion}

\subsection{Errors and Biases}\label{errors-and-biases}

When compared to major polling companies and their ongoing polls
published in the media, it seems that several parties are
under-represented while others are over-represented, even after
introducing our statistical controls. Some notable examples are:

\begin{itemize}
\itemsep1pt\parskip0pt\parsep0pt
\item
  The left-wing party Meretz gets 9 seats but only 4-5 seats in most
  polls.
\item
  The right-wing party Yachad gets 9-10 seats but only 4-5 in most
  polls.
\item
  The separadic-orthodox party Shas gets only 4-5 seats but 8 in most
  polls.
\item
  The right-wing party Israel Beytenu gets only 0 seats but 4-5 in most
  polls.
\end{itemize}

One possible source of bias is the influence of abstention (non-voting).
Our approach doesn't include a mechanism to assess changes in turn-out.
While we did ask our respondents whether they abstained in 2013, it
would be naïve to assume that they represent the non-voting population.
A non-voter is presumably indifferent and would not participate in our
poll. Those who do participate, probably intend to vote in 2015. We
could therefore easily reach the false conclusion that turn-out will
increase to nearly 100\% giving those users who reported abstention in
2013 unreasonably high weights. Eventually we chose to ignore possible
changes in abstention, implicitly assuming that the voting population is
constant and that we need only to infer if and how they will change
their vote. In particular, there are media reports that turn-out will
increase dramatically in the Arab population, which could increase the
number of seats for the Arab Union.

Other possible sources of bias are several demographic variables that we
did not control. Since we are already control for the 2013 election
results (\emph{i.e.} our weighted sample agrees with the official 2013
results), the following question is of relevance to the sources of bias:
In what way are our respondents that voted for party \(i\) in 2013
different from the actual voting population that voted for party \(i\)
in 2013? Some variables that may have played a role in biasing our
sample could be:

\begin{itemize}
\itemsep1pt\parskip0pt\parsep0pt
\item
  Age. It is reasonable to assume that young voters are more likely to
  vote for new, small, niche or extreme parties than are older voters.
  For example, older voters who voted for Ha'Likud in 2013 are more
  likely to vote again to Ha'Likud than are youngers voters, who are in
  turn more likely to switch to Yachad or Kulanu. This same bias could
  explain the high number of seats projected for Merez and the low
  number of seats projected for Israel Beytenu.
\item
  Sex. It is possible that men changed their minds (between 2013 and
  2015) in a manner different from women. Ha'Mahane Ha'Zioni has a high
  level of female representation, including Livni who is set to become
  prime minister through rotation with Herzog. As an example, women who
  voted for Yesh Atid in 2013 might be more likely to switch to
  Ha'Mahane Ha'Zioni than men who voted for Yesh Atid in 2013.
\end{itemize}

Another source of error could be the sample size. 2013 votes were only
collected from respondents after February 13th, and only in Android
devices. Therefore, the 2013 dataset contains only
\textasciitilde{}2,400 votes, roughly a third of our entire sample. Due
to our methodology, it is imperative to have a reasonable sample size
for each voter `weight class', which is decided in our case by the
voter's 2013 vote. For instance, we might have 400 respondents who voted
Ha'Likud in 2013, but only 5 that voted Shas. Those few respondents in
the 2013-Shas weight class will get high weights, likely leading to
large errors. Due to time limitations, we did not have the time to
estimate these errors, so it is likely that they explain at least some
of the deviance in our forecast. In particular, this could explain the
low number of seats projected for Shas, Yahadut Hatora or Israel
Beytenu.

\subsection{Conclusion}\label{conclusion}

Ha'Midgam app offered Israelis a chance to express their voice in an
online, realtime, open poll and to view a live forecast of the upcoming
2015 Israeli elections results. It is likely that our poll suffers from
sample bias. However it serves as an important proof of concept. We
believe that with better bias controls, additional demographic
information and a marketing effort targeted at specific
under-represented demographics a smartphone app can become a precise
poll and make forecasts as good as the major national polls.

\end{document}